%% file: 0.main.tex
\documentclass[acmsmall, review=false]{acmart} 

\AtBeginDocument{%
  \providecommand\BibTeX{{%
    \normalfont B\kern-0.5em{\scshape i\kern-0.25em b}\kern-0.8em\TeX}}}

\setcopyright{acmcopyright}
\copyrightyear{2018}
\acmYear{2018}
\acmDOI{10.1145/1122445.1122456}

\usepackage{booktabs} 
\usepackage{url}

\usepackage{caption}
\usepackage[all]{nowidow}
\usepackage{wrapfig}
\usepackage{array}
\usepackage{arydshln}
\usepackage{tabularx}
\usepackage{multirow}
\usepackage{arydshln}
\newcolumntype{L}[1]{>{\raggedright\let\newline\\\arraybackslash\hspace{0pt}}m{#1}}
\newcolumntype{C}[1]{>{\centering\let\newline\\\arraybackslash\hspace{0pt}}m{#1}}
\newcolumntype{R}[1]{>{\raggedleft\let\newline\\\arraybackslash\hspace{0pt}}m{#1}}

\usepackage{wrapfig,lipsum,booktabs} 
\usepackage{enumitem}
\usepackage{float}       
\usepackage{caption}
\usepackage{subcaption}
\usepackage{tabularx}    
\def\authnotes{1}
\newcounter{notectr}[section]
\newcommand{\thenote}{\thesubsection.\arabic{notectr}\refstepcounter{notectr}}

\newcommand{\note}[2]{$\ll$#1~\thenote: #2$\gg$}
\newcommand{\cnote}[1]{\ifnum\authnotes=1 \textcolor{blue}{\note{Comment:}{#1}}\fi}

\usepackage{graphicx}
\usepackage{subcaption}

\usepackage{longtable}

\acmYear{2024}
\copyrightyear{2024}
\acmConference[CSCW '26]{TBD}
\acmBooktitle{CSCW}
\acmDOI{TBD}
\acmISBN{TBD}

\begin{document}



\title[Reproductive Well-being \& ReWa]{From Literature to \textit{`ReWA'}: Discussing Reproductive Well-being in HCI}



\author{Hafsah Mahzabin Chowdhury}
\affiliation{
  \institution{University of Illinois Urbana-Champaign}
  \city{Urbana}
  \state{Illinois}
  \country{USA}}
\email{hafsahc2@illinois.edu}

\author{Sharifa Sultana}
\affiliation{
  \institution{University of Illinois Urbana-Champaign}
  \city{Urbana}
  \state{Illinois}
  \country{USA}}
\email{sharifas@illinois.edu}

\begin{abstract}
Reproductive well-being is shaped by intersecting cultural, religious, gendered, and political contexts, yet current technologies often reflect narrow, Western-centric assumptions. In this literature review, we synthesize findings from 147 peer-reviewed papers published between 2015 and 2025 across HCI, CSCW and social computing, ICTD, digital and public health, and AI for well-being scholarship to map the evolving reproductive well-being landscape. We identify three thematic waves that focused on early access and education, cultural sensitivity and privacy, and AI integration with policy-aware design, and highlight how technologies support or constrain diverse reproductive experiences. Our analysis reveals critical gaps in inclusivity, with persistent exclusions of men and non-binary users, migrants, and users in the Global South. Additionally, we surfaced the significant absence of literature on the role of stakeholders (e.g., husband and family members, household maids and cleaning helping hands, midwife, etc.) in the reproductive well-being space. Drawing on the findings from the literature, we propose the \textit{ReWA} framework to support \textit{reproductive well-being for all} agenda through six design orientations associated with: location, culture, and history; polyvocality and agency; rationality, temporality, distributive roles, and methodology. 

\end{abstract}







\settopmatter{printfolios=true}

\maketitle

\input{1.intro}
\input{2.lit}
\input{3.methods}

\input{4.find}

\input{5.insights}

\input{6.discussion}
\input{6.lim-fw-con}



\bibliographystyle{ACM-Reference-Format}
\bibliography{0.main}

\end{document}

%% file: 1.intro.tex
\section{Introduction}
The landscape of reproductive well-being is shaped by cultural, religious, gendered, and political contexts, making it a highly sensitive and often stigmatized domain \cite{3, 5, 27, 63, 76, 77, 108, 122, 135, 140}. Across different societies, topics such as menstruation, infertility, pregnancy loss, abortion, menopause, and men's reproductive well-being are navigated through varying norms of silence, shame, or openness, with certain populations, such as men, migrants, religious minorities, low-income communities, and non-binary individuals, remaining chronically underrepresented in both healthcare and research \cite{3, 13, 14, 17, 31, 32, 33, 34, 35, 47, 64, 113, 122, 131}. In response to gaps in access and education, a range of digital technologies have been developed, including menstrual tracking apps, fertility predictors, pregnancy monitoring systems, and AI-driven reproductive well-being tools \cite{3, 5, 23, 27, 49, 62, 76, 77, 140, 148}.


However, the design space of many of these technologies is dominated by Western ideas about health and well-being, leading to one-size-fits-all designs that do not meet the different needs, perspectives, and experiences of people from diverse cultures and backgrounds \cite{5, 6, 18, 19, 27, 30, 42}. Consequently, users in the Global South, migrant populations, and marginalized gender groups frequently find that available tools do not reflect their realities or reproductive priorities \cite{5, 27, 42, 92}. Privacy concerns further complicate the use of digital health tools, with studies showing that users often feel insecure and unsafe about how sensitive reproductive data are collected, shared, and potentially weaponized against them, particularly after major legal shifts like the overturning of Roe v. Wade in 2022 \cite{24, 30, 43, 45, 53, 125}. Note that Roe v. Wade was a landmark decision by the United States Supreme Court in 1973 that established the constitutional right to abortion \cite{roevwade1973}. The Court ruled that a woman's right to choose to have an abortion was protected under the right to privacy implied by the 14th Amendment of the U.S. Constitution. This decision effectively legalized abortion nationwide, preventing states from banning it in the early stages of pregnancy.


Beyond women-centered designs, there is a growing recognition that men's reproductive well-being needs are critically underexplored, with a few studies documenting how male infertility, emotional distress, and paternal engagement remain largely invisible across both digital platforms and clinical systems \cite{31, 32, 33, 34, 35}. Even with this small amount of literature on men's reproductive well-being \cite{31, 32, 33, 34, 35}, the domain primarily suffers from this binary gender division of humans in reproductive space and misses out the roles of other stakeholders such as husbands and family members, household maids, cleaning helping hands, midwives, etc. Altogether, these gaps leave reproductive well-being research and technologies fragmented, with many populations, experiences, and vulnerabilities still overlooked. To make reproductive well-being more holistically supportive for everyone, it is crucial to locate the gaps in existing literature.   


We look to understanding how the domain of reproductive well-being is shaped, supported, and sometimes constrained by digital technologies, design practices, and sociotechnical systems, as well as the significant gaps in this scholarship. We examined research papers published in the past ten years in human-computer interaction (HCI), computer-supported cooperative work (CSCW), digital well-being, information and communication technology and development (ICTD), and AI for well-being, which helped us connect diverse perspectives and construct a more comprehensive picture of the reproductive well-being landscape. We started with more than 300 papers published between 2015 and 2025 on diverse reproductive experiences, including menstruation, fertility, abortion, pregnancy, maternal care, menopause, and men's reproductive well-being. From them, we thoroughly reviewed 147 papers that directly addressed the topics of interest and mapped the major concerns, methodological approaches, design interventions, and ethical challenges associated with reproductive well-being. 

The findings from reviewing the literature highlight common themes, ongoing challenges, and gaps in the literature. While topics like stigma, cultural mismatch, data privacy, and design justice appeared in discussion, they are still inconsistently addressed in design and practice. Significant gaps in theory, research, design, and practice around the noticeable absence of other stakeholders, missing contextual mapping across geographies, and poor recognition of alternative temporalities were identified. Drawing on the findings from the literature, we propose the \textit{reproductive well-being for all (ReWA)} framework to support more holistic reproductive care and well-being for all through six design orientations associated with: location, culture, and history; polyvocality and agency; rationality, temporality, distributive roles, and methodology. This framework will work as a set of practical guidelines to design fair, accountable, and sustainable infrastructure and computing tools to support reproductive well-being for all, as well as a critical lens to analyze whether an existing infrastructure or tool is supportive of reproductive well-being for all.

This paper makes four key contributions to the field of reproductive well-being and technology. \textbf{First,} our literature review maps how reproductive well-being experiences and outcomes have been shaped by the interplay of cultural contexts, infrastructural constraints, sociotechnical systems, and user agency in the past ten years. \textbf{Second}, we outline the challenges, design problems, and ethical issues found in digital tools and systems related to reproductive well-being in the past ten years' research and show them in a thematic timeline. \textbf{Third}, we identify three major gaps in current theory, research, design, and practice: noticeable absence of other stakeholders, missing contextual mapping across geographies, and poor recognition of alternative temporalities. \textbf{Fourth}, to lead the domain toward reproductive well-being for all agenda, we drawing on our review and we propose the \textit{Reproductive Well-being for All (ReWA)} framework, a critical and practical guide structured around six design orientations (location, culture, history; polyvocality and agency; rationality; temporality; distributive roles; and methodology) to support the creation and evaluation of fair, accountable, and inclusive reproductive technologies

Note: Not all studies reviewed in the manuscript focus exclusively on digital technologies. While many papers examine apps, AI tools, and digital health interventions, others explore broader social, cultural, and infrastructural factors that shape reproductive well-being experiences. We include both technological and non-technological research in this review, recognizing that digital systems are interconnected with broader socio-technical and cultural realities.

%% file: 2.lit.tex
\section{Related Work}
Literature review paper provides a broad overview of how a field develops and where important gaps remain. For example, Dell and Kumar reviewed human-computer interaction for development research published between 2009 and 2014, and showed how the field expanded into new regions and topics while identifying challenges in collaboration between researchers and practitioners \cite{insandouts}. Another review of migration and displacement in human-centered computing by Sabie et al. highlighted how the focus shifted from immediate needs toward deeper political and emotional dimensions \cite{migration}. Similarly,  a systematic review of social computing and social media research from 2008 to 2020 talked about common methods, but also pointed out gaps such as the limited use of feminist approaches \cite{shibuya2022mapping}. These reviews show the value of mapping existing research to surface blind spots and move fields toward more responsible and inclusive practices.

\subsection{Review of Reproductive Well-being Literature in HCI and Social Computing}
In the reproductive well-being field, systematic reviews have revealed how technologies support or fail to support diverse reproductive needs. For example, Ibrahim et al. conducted systematic reviews on women's reproductive well-being research in HCI, highlighting persistent gaps in demographic reporting, participant diversity, and methodological transparency \cite{8}. Similarly, Navarro and Prabhakar reviewed digital interventions for women's reproductive well-being in Latin America and emphasized the urgent need for culturally situated reproductive well-being technologies in such regions where challenges like low literacy, stigma, and infrastructural gaps shape how digital interventions are received and used \cite{27}.  Additionally, recent reviews on digital health tools, such as chatbots for sexual health, emphasize how these technologies tend to privilege English-speaking, high-income populations, leaving low-income and culturally diverse users underserved \cite{139}. Gagnon and Redden investigated reproductive well-being outcomes among migrant women in Western countries and found major gaps in research beyond pregnancy-related care,  such as menopause, sexual health behaviors, and sexually transmitted infections (STIs). Their review also emphasized the lack of methodological consistency and the need for better migration-related data collection practices \cite{112}. These findings highlight how even within reproductive well-being research, significant blind spots persist, especially concerning underrepresented populations.

Inspired by previous literature reviews that explore the evolving fields while uncovering gaps, we undertake a similar synthesis at the intersection of reproductive well-being and HCI. While there are existing reviews focused on specific areas, such as migrant women’s reproductive outcomes \cite{112} and Latin American reproductive well-being technologies \cite{27}, to our knowledge, no prior work brings together stigma, cultural mismatch, gender inclusion, privacy concerns, and technological design challenges across the broader reproductive well-being landscape.  By bringing together research from HCI, CSCW, ICTD, AI ethics, and public health, this review offers a more holistic understanding of how reproductive well-being is shaped, supported, and sometimes constrained by digital and socio-technical interventions. 

\subsection{Understudied Gaps in Reproductive Well-being Research}
Over the past decade, increasing research has examined reproductive well-being through the lenses of design, technology, and health equity \cite{3, 5, 76, 108}. These studies have addressed a range of topics, including menstrual tracking, fertility planning, maternal care, and emotional support during menopause \cite{3, 23, 62, 77, 82}. A number of studies have also called for more inclusive systems that recognize the needs of users beyond menstruating individuals, such as LGBTQ+ people, non-binary users, and those experiencing infertility or pregnancy loss \cite{63, 82, 92, 108}. Yet, despite these efforts, significant gaps remain in both research and real-world application.

Our review shows that frequently mentioned topics like inclusivity and ethical data practices are often underexplored, sidelined, or not meaningfully translated into real-world design. In many cases, these conversations remain aspirational: papers suggest new directions, but there is little evidence that designers, developers, or platforms have adopted them. For example, the need to move beyond binary gender frameworks and conception-focused fertility tools has been emphasized \cite{63, 92, 108}. However, such recommendations have had limited impact on mainstream reproductive technologies. These blind spots themselves remain understudied as a pattern worth investigation.

To address this gap, we conducted a decade-long mapping of research on reproductive well-being and examined what has been covered, what remains overlooked, and where accountability continues to fall short. By looking at different ideas, missing topics, and lack of progress, this review highlights the gap between what research calls for and what actually gets done in design. This mapping is not just about summarizing what exists, but about showing what continues to be missing, and why those absences matter.

%% file: 3.methods.tex
\section{MAPPING THE LITERATURE: Method and Strategy}
\input{3.tab-1}

This section outlines our approach to reviewing literature at the intersection of reproductive health and interactive computing, focusing on research published between 2015 and 2025. We began with an initial pool of over 274 papers and narrowed this down to 147 for detailed analysis. We aimed to evaluate how existing technologies shape, support, or act as a barrier in terms of reproductive health, along with how reproductive health is experienced, perceived, and mediated across different gender identities and social contexts.

Our structured literature survey searched through academic databases including ACM Digital Library, IEEE Xplore, Scopus, and Google Scholar. We curated this list based on (1) top-tier peer-reviewed conferences and journals in Human-Computer Interaction, (2) special tracks and workshops addressing reproductive health and digital systems, and (3) research explicitly focused on reproductive health, sexuality, fertility, menstruation, and care systems across socio-technical contexts. The search terms included, Reproductive Health and HCI," "Menstrual Tracking," "Fertility self-tracking," "Maternal health technology," "Reproductive justice," "Sexual health education and stigma," "Men’s reproductive health," "Male infertility," "Male sexual health technologies", which ensured an extensive coverage of the topic. Table \ref{tab:param} summarizes the search parameters of our literature review.

We included papers from conferences, journals, and workshops such as, CHI, CSCW, DIS, NordiCHI, IDC, MuC, CLIHC, AfriCHI, COMPASS, ICTD, CHI PLAY, SAICSIT, CHI4Good, CHItaly, TEI, ACM Transactions on Computer-Human Interaction (TOCHI), PACMHCI, JMIR (MHealth and UHealth), IEEE Transactions on Technology and Society, Proceedings on Privacy Enhancing Technologies (PoPETs), and ACM Journal of Responsible Computing. We used a rubric that captured the key dimensions, such as, "when","where","who","why",and "how", to systematically organize and analyze the literature. Definitions for each of these categories are provided in Table \ref{tab:review-rubric} All papers were coded according to this framework.

\input{3.tab-2}

%% file: 3.tab-1.tex
\begin{table}[b!]
\centering
\begin{tabular}{|l|p{9cm}|}
\hline
\textbf{Date filter} & 2015 $\leq$ year $\leq$ 2025 \\
\hline
\textbf{Search fields} & Title, abstract, keyword, and full-text body \\
\hline
\textbf{Quality Assessment} & Peer-reviewed long and short research articles, manually screened for relevance and quality. \\
\hline
\textbf{Search terms} & 
"Reproductive Health and HCI," "Menstrual Tracking," "Fertility self-tracking," "Maternal health technology,"  "Reproductive justice," "Sexual health education and stigma,"  
"Men’s reproductive health," "Male infertility,"  "Male sexual health technologies" \\
\hline
\textbf{Date of Query} & Initial: January 2025; Updates: March 2025 and April 2025 \\
\hline
\end{tabular}
\caption{Search parameters used for identifying literature on reproductive health and HCI}
\label{tab:param}
\end{table}

%% file: 3.tab-2.tex
\begin{table}[!b]
\centering
\begin{tabular}{|l|p{11cm}|}
\hline
\textbf{Criteria} & \textbf{Definition} \\
\hline
When & The timeline of publication, capturing evolving research from 2015 to 2025 across three key waves of development in reproductive health technologies. \\
\hline
Where & The geographical focus of each study, highlighting regional patterns: from the United States to underrepresented contexts like Sub-Saharan Africa, Latin America, and South Asia. \\
\hline
Who & The target population addressed, including women, men, adolescents, LGBTQ+ individuals, migrants, Muslim users, and healthcare providers. \\
\hline
What & The key challenges and themes addressed across reproductive health research, ranging from inclusive access and privacy concerns to AI ethics, and systemic marginalization. \\
\hline
Why & The goal or motivation behind the study, such as addressing stigma, improving access, supporting emotional wellbeing, promoting education, or navigating reproductive rights. \\
\hline
How & The technologies and research methods used, including apps, wearables, chatbots, social media, surveys, participatory design, speculative design, ethnographic studies, or data/privacy audits. \\
\hline
\end{tabular}
\caption{Criteria used to organize and analyze the reproductive health literature in HCI. Definitions for "when," "where," "who," "what," "why," and "how" capture key dimensions used in our review to identify trends, gaps, and research directions.}
\label{tab:review-rubric}
\end{table}

%% file: 4.find.tex
\section{Literature Comprehension}
This section overviews the past ten years' reproductive health in HCI scholarship and comprehend \textbf{\textit{when}} and \textbf{\textit{where}} this work has focused, \textbf{\textit{who}} the target users were, \textbf{\textit{what}} technologies and interfaces were designed or studied, \textbf{\textit{why}} these systems were developed, and \textbf{\textit{how}} reproductive experiences were conceptualized and represented. Due to the volume of contributions in this evolving domain, we include one or two representative examples for each major theme. We deliver a focused, high-level summary below that (1) allows readers to understand the current landscape of reproductive health research in HCI, and (2) brings awareness to the expanding range of themes and approaches, emphasizing both the areas that have received concentrated attention and those that remain underrepresented. 

\subsection{When: 2015 to 2025}
The studies we have reviewed on reproductive health and HCI started from 2015 to 2025. This review captures the evolution of reproductive health research from early interventions focused on basic education and access in low-resource settings to addressing AI ethics, privacy, and intersectional justice in reproductive health technologies. Figure \ref represents the number of papers published each year during our review period, which reflects the field's growing complexity and diversification in both technological approaches and population focus. 

\subsubsection{First Wave: Access and Early Community Building} Between 2015 and 2017, early research in the intersection of reproductive health and HCI focused on expanding basic access to health education and creating initial spaces for community support. Four studies focused on low-cost, accessible technologies and targeted basic access issues through mobile health (mHealth) initiatives, particularly in low-resource settings such as India, Uganda, and Pakistan. For example, SMS interventions to support maternal health education \cite{73,74}, along with using projectors to facilitate group maternal health education sessions \cite{72}. Additionally, early mobile applications were used to address specific needs, such as a web-based mobile app that provided pregnancy-related information in English and Vietnamese \cite{110}, while another application monitored perinatal depression among low-income mothers \cite{87}. Additionally, attempts were taken through games to engage adolescents in menstrual and reproductive health learning through interactive simulations \cite{104}.

In Western contexts during this period, 13 studies explored how early social platforms could support reproductive well-being communities. For example, forums provided trimester-specific peer mentoring \cite{107}, Facebook groups supported postpartum mental health \cite{90}. Meanwhile, physical health technologies were also introduced, which reflected parallel interest in improving embodied care practices, such as the Labella, a wearable technology system designed to help women learn about their intimate anatomy \cite{25}.

This early wave of research set the foundation for future developments by focusing on making health care more accessible, affordable, and community-centered. Technology was used not as a high-tech solution, but as a basic tool to share important health information and build early networks of peer support. Such research concerns and topics were dominant before 2017.

\subsubsection{Second Wave: Cultural Sensitivity, Diversification, and Privacy Awareness} Between 2018-2020, HCI and reproductive health research entered a second wave marked by technological diversification, cultural sensitivity, and growing concern around data privacy. During this period, researchers started expanding their scope beyond foundational health information delivery tools to address more complex reproductive health needs. Studies increasingly started analyzing community-based digital spaces, including Reddit, Facebook groups, and Whatsapp \cite{66, 105, 106, 109, 111}. These spaces became important for intimate storytelling and support for pregnancy loss, menopause, and postpartum care. Whatsapp was used as a peer education platform for taboo topics in low-resource settings \cite{66, 105, 106, 109, 111}. 

During this period, menstrual health technologies became more complex, with growing attention on how these applications handled users' private data. For example, \textit{Formoonsa Cup} (a reusable menstrual cup) brought together menstrual health innovation and civic activism to change laws and break taboos, using crowdfunding and online petitions in East Asia \cite{70, 132}. Meanwhile, studies reviewing menstrual tracking apps showed that many platforms had vague privacy policies and weak consent protections \cite{56}. Hence, along with this diversification the critiques of reproductive surveillance grew. Papers began interrogating the data flows within menstrual and fertility tracking apps and devices. Studies uncovered opaque third-party data sharing of these apps, and the rising anxiety among the users over surveillance in post-Roe digital environments \cite{56, 57}. 

\begin{figure}[!t]
    \centering
    \includegraphics[width=1\textwidth]{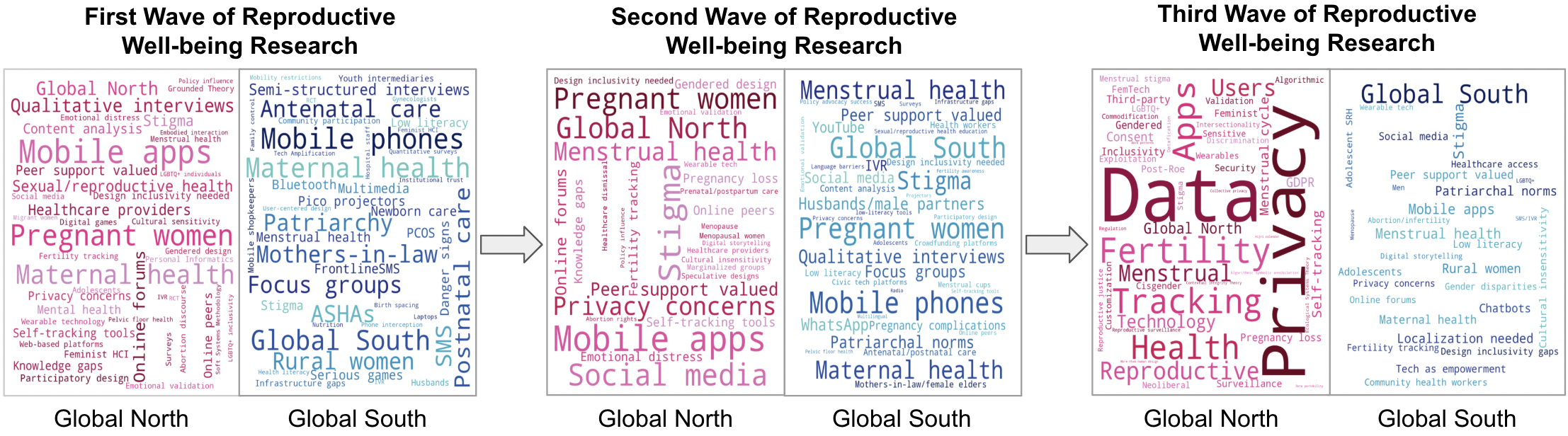}  
    \caption{Thematic timeline of the reproductive well-being scholarship, the word clouds represent the key focus areas of research, design, and practice in each wave in both the Global North and the Global South.}
    \label{fig:waves}
\end{figure}

While the second wave introduced more critical conversations about privacy and ethics in Western contexts, basic access to reproductive health support remained an ongoing challenge in many parts of the global south, much like the first wave. Initiatives like vTaiwan and low-cost SMS and Interactive Voice Response (IVR) Technology continued to prioritize expanding reach over advancing debates around data rights \cite{70, 71, 80}. Such research concerns and topics were dominant before 2020.

\input{3.tab-3.tex}

\subsubsection{Third Wave: AI Integration, Privacy, and Policy-Aware Design} Between 2021 and 2025, research in this field rapidly expanded, became more critical, and covered a wide range of topics. A large portion of the papers focused on how AI, privacy, misinformation, and policy shifts shaped the future of reproductive health technologies. Studies responded directly to socio-political events, namely, the overturning of Roe v. Wade, expressing concerns about surveillance, algorithmic bias, and data security to the center of design discussions \cite{24, 43, 45, 53, 118, 119, 150}. 

AI integration became prominent in this phase, but with different goals across contexts. In the global north, AI-integrated menstrual health and fertility apps drew critical attention to issues of transparency and bias in algorithmic predictions \cite{15, 26, 48, 52, 55, 57}. Researchers responded to these challenges by addressing the need for explainability in algorithmic tools, participatory auditing, and stronger post-Roe privacy protections \cite{53, 61}. 

Privacy and policy awareness became central themes. Studies reported users' adoption of VPNs and privacy workarounds, researchers critiquing FemTech apps for inadequate consent mechanisms and third-party data sharing  \cite{42, 61, 120, 125, 150}. Misinformation became another area of concern pushing the development of fact-checking systems, and NLP-based discourse analysis on social media platforms like Reddit and Twitter \cite{24, 117, 126}. 

Consequently, the third wave also highlighted how many mainstream technologies still failed to meet the needs of marginalized users. Studies revealed that existing menstrual health tracking apps were often built around the assumptions of regular, cisgender female users, excluding those with non-normative cycles, gender identities, or medical conditions. Users expressed the need for contextualized symptom insights, flexible data-sharing options, and more inclusive educational resources \cite{67}. These unmet needs reflected design gaps and the lack of culturally responsive and emotionally coordinated technologies. 

In contrast, researchers in the Global South focused on foundational access and education. Chatbots like \textit{SnehAI}, \textit{MAI}, and \textit{Raaji} showed how these tools could expand access to reproductive care \cite{5, 49, 140}. Hence, they were designed to deliver reproductive health education through culturally sensitive, language-specific, and low-bandwidth platforms. These systems prioritized reach, inclusion, and stigma reduction. At the same time, some studies focused on addressing infrastructural gaps. Researchers from low-resource settings built low-cost tools such as wearable sanitary sensors, SMS-based training apps, and low-bandwidth health interfaces and underscored that their fair access to technology is still a major challenge \cite{79, 88, 108, 135}. 

These developments suggest that while some regions moved towards algorithmic accountability and policy innovation, others were still grappling with basic digital inclusion and literacy. Hence, this third wave reflects a maturing but uneven landscape. Researchers emphasized that innovation alone is not enough, technologies for reproductive care must be safe, inclusive, and designed to reflect diverse needs and histories of exclusion.

\subsection{Where: Geographical Distribution}

\begin{figure}[!t]
    \centering
    \includegraphics[width=1.0\textwidth]{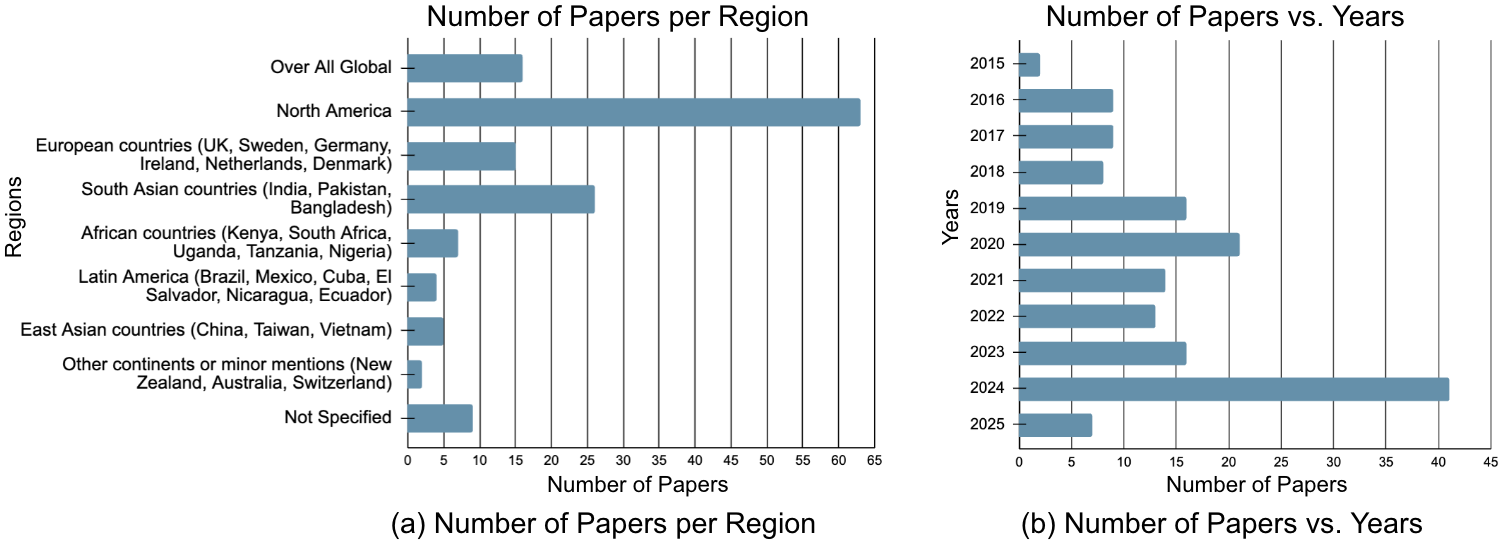}  
    \caption{Distribution of total number of papers (n=147): (a) number of papers per region, and (b) number of papers vs. year.}
    \label{fig:fig2}
\end{figure}

Research in reproductive health and HCI over the past decade has been unevenly distributed across global regions. As demonstrated in Figure \ref{fig:aligned_combined}(b) , the United States is dominating with over 60 papers, contributing to nearly half of the overall body of work, while regions like Africa, Latin America, and the global south remain under-represented across the First Wave (2015–2017), Second Wave (2018–2020), and Third Wave (2021–2025).

The United States is the most represented region with 41\% of the overall reviewed literature. In the First Wave, US-based studies focused on self-tracking tools and community support through online platforms \cite{16, 90, 107}. The second wave saw diversification of reproductive health tools, which includes early innovations for menopause and postpartum care, as well as the emergence of privacy concerns in popular menstrual tracking apps \cite{28, 84, 56}. By the third wave,  the focus shifted to AI integration with tools like chatbots for reproductive health education, and policy-aware designs addressing post-Roe privacy concerns, such as VPN usage and surveillance critiques \cite{119, 150}. This shift puts the US ahead in AI-integrated and policy-aware reproductive health technologies, moving faster than other regions in both tech and legal response. European countries being moderately represented, also focus on privacy in fertility apps, ethical critiques of FemTech, and tools for LGBTQ+ and marginalized users. 

India follows as the most represented country in the Global South with 11\% of the reviewed papers. This demonstrates a growing focus on context-specific reproductive health technological solutions. During the first wave, India prioritized low-tech maternal health educational interventions \cite{72}. The second wave emphasized culturally sensitive tools such as comics for menstrual health education, and social media to share health information \cite{17, 109}. In the third wave, India adopted AI-driven chatbots for educational purposes, along with exploring sensor-based wearables \cite{5, 36, 62}. Regions like Pakistan and Bangladesh also show localized AI solutions but have minimal attention to data governance or regulatory frameworks. South Asia's development from low-tech to AI-integrated reproductive health technological solutions reflects advancement, though it lags behind the US in privacy infrastructure and policy-focused designs.

Africa, Latin America, and the Middle East remain significantly underrepresented, focusing on low-tech interventions. In the first wave, their contributions center on SMS-based interventions to provide for maternal education and alerts \cite{71, 73}. The second wave continued this trend with bidirectional SMS in rural Kenya \cite{41}. The third wave saw a slight shift with mobile apps for rural Nigeria and digital storytelling for maternal wellbeing in South Africa \cite{77, 82}. Africa's consistent reliance on low-tech solutions highlights a persistent infrastructural gap, along with a lack of advancement in AI, privacy, or policy-focused reproductive health technology, leaving it far behind regions like the US and India.

Other regions, such as Latin America, Australia/New Zealand, and Arab countries, are similarly underrepresented \cite{27, 58, 110}. These regions remain focused on foundational health education and tracking, missing the AI-integrated and policy-aware advancements seen in the US, which reflects a global digital divide. 

While the Global North rapidly moved towards sophisticated design, algorithmic transparency, and policy-aware tools, many Global South regions remained focused on access, inclusion, and basic education, which underscores a persistent disparity in technological advancement and research investment across geographies.

\subsection{What: Challenges and Dynamics in Reproductive Health Technologies}

Studies in this field have not only introduced a wide range of technologies, but also revealed the recurring challenges that shape their impact. Across the reviewed papers, three major themes emerged: (1) Cultural Inclusivity and Accessibility, (3) Structural Exclusions and Underrepresented Populations, and (3) Privacy and Ethical concerns. 

\subsubsection{Cultural Inclusivity and Accessibility}
A major challenge in the development of reproductive health technology is ensuring cultural inclusivity and accessibility, particularly underserved and culturally diverse populations who often face barriers to effective health technology use. 
    
In low-resource settings, prenatal and postnatal education is delivered through SMS platforms, but the reliance on English limited accessibility for non-literate or non-English-speaking women, highlighting a language barrier \cite{71, 73}. Even after bringing innovations in these areas, such as educational comics or offline educational workshops to provide reproductive health education in a culturally sensitive manner, these solutions struggled to reach women without smartphone access. 
    
At the same time, even in high-tech environments, users whose experiences do not conform to the "norm" often find themselves excluded. Many menstrual health tracking apps are built for users with predictable cycles and cisgender identities. This caused frustration among users as their lived experiences, such as irregular periods, menopause transitions, or gender dysphoria were not acknowledged or supported \cite{67}. 
    
Together, these studies underscore that cultural inclusivity is not just about localization, it's about designing technologies that are context-aware, emotionally responsive, and capable of supporting users across diverse life stages, literacies, and identities.

\subsubsection{Structural Exclusions and Underrepresented Populations}

Despite a decade of growth in this field, many populations remain marginalized in both design and research. One consistent gap is the limited representation of men, non-binary and LGBTQ+ individuals, and other marginalized populations. Only a few studies address men's reproductive well-being, such as Trak: Sperm Health for fertility tracking in the US, and AI chatbots for men's health \cite{31, 33}. Non-binary and LGBTQ+ individuals are similarly excluded, as most technologies assume a binary gender framework \cite{26, 138}. Additionally, while several apps claim to serve a "global" user base, design and data training are often West-centric, reflecting assumptions rooted in White, educated, cisgender, and urban populations \cite{1, 19, 56, 149}. Hence, these apps fail to serve marginalized users, where app illiteracy, cultural misalignment, and lack of localized support persist.

\subsubsection{Privacy and Ethical Concerns}

As reproductive health technologies became more data-driven, privacy and ethics became a central concern, particularly in sensitive contexts such as abortion regulation, surveillance capitalism, and algorithmic opacity. In the Second Wave (2018–2020), menstrual health trackers were found to share users' data with third-party trackers often without consent \cite{56, 57}. By the Third Wave (2021–2025), the overturning of Roe v. Wade in the US amplified privacy concerns, and users were concerned about data being misused to incriminate them in abortion-sensitive regions, raising the usage of privacy tools like VPNs \cite{43, 150}. AI-driven tools introduced additional ethical challenges, for instance, raising concerns about algorithmic bias and transparency, particularly when applied globally without local context \cite{119}.
    
Notably, these concerns were not equally distributed as in the Global South, access often came before privacy. Studies from those regions showed that users mainly needed trustworthy health information and local support \cite{41, 64, 77, 110}. In such contexts, the ethical imperative was not just to secure data but to ensure that protections did not hamper usability or outreach.

\subsection{Who: Target Populations}
\begin{wrapfigure}{r}{0.55\textwidth}
\vspace{-20pt}
\includegraphics[width=0.99\linewidth]{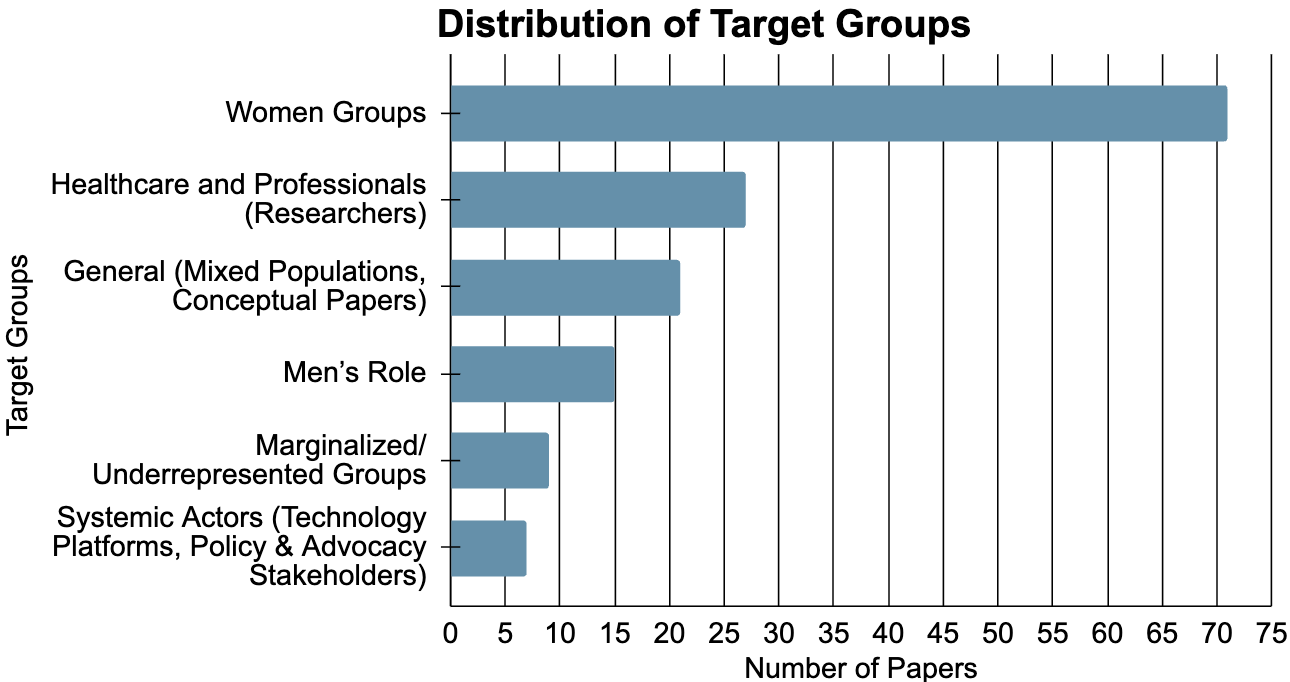} 
\caption{Distribution of target groups: women, men, professionals, and other stakeholders.}
\label{fig:wrapfig}
\vspace{-10pt}
\end{wrapfigure}


The reviewed papers highlight diverse participants engaged in or impacted by reproductive health and technology interventions. These include women across life stages, healthcare providers, researchers, marginalized groups, adolescents, men, and various institutional actors. Fig.3 visualizes the distribution of these populations across the 147 papers analyzed. The following subsections provide a detailed breakdown of how the groups' presence and role were addressed in the literature. 

\subsubsection{Women and Menstruating Individuals} A dominant focus of the literature examines menstruating individuals in around 34 papers. These include individuals who uses period tracking or fertility applications, suffers from health conditions, such as, PCOS or endometriosis \cite{19,23,67,100}. These studies reveal disparities in menstrual education, where 20-30 year olds from educated backgrounds are overrepresented \cite{67, 100}, while marginalized groups face data gaps \cite{144}. For example, Muslim women navigating religious practices during menstruation \cite{7, 147}. Additionally, a smaller subset explores menstruators with health conditions, highlighting how apps lack inclusivity for these health conditions, such as PCOS, while failing to accomodate cycle irregularities or pain management needs \cite{14, 100}. Despite menopause being a major reproductive milestone, women experiencing menopause remain underrepresented \cite{27, 28, 105}. 

\subsubsection{Pregnant/Postpartum Women} Twenty-two papers focus on maternal health  (e.g., \cite{73, 74, 87}), with 8 specifically studying postpartum experiences \cite{81, 90}. Some studies address high-risk pregnancy complications to understand  sociocultural and infrastructural barriers that affect access to opportunities for digital health support, such as, Indigenous Kichwa women \cite{79}, 
African-American women with perinatal depression \cite{81, 102}, Anemic pregnant women in rural India \cite{78}. These studies highlight systemic gaps in care, particularly for marginalized groups facing intersectional challenges \cite{9, 96}.

\subsubsection{Healthcare Providers} From midwives (crucial for community care in low-resource settings) to Accredited Social Health Activists (ASHAs), fertility specialists, and general practitioners, healthcare providers are crucial in bridging reproductive technologies with practice \cite{72, 88, 4}. Studies critique the lack of provider training on tech interfaces \cite{110} and cultural competencies \cite{79}. 

\subsubsection{Adolescents} Reproductive health of adolescents are discussed in 8 papers (e.g, \cite{12, 104}), one of them revealing the digital access disparities where a striking 93\% male user base was found in India's \textit{SnehAI} sexual health chatbot, reflecting cultural barriers preventing girls from accessing digital SRH resources \cite{5}. Additionally, urban, english-speaking adolescents are represented in multiple studies while rural youth remains underrepresented \cite{5, 17}. Studies also introduced innovative interventions, such as, multiplayer game to increase reproductive health knowledge among adolescent users \cite{12}.

\subsubsection{Men} Men remain minimally represented (15 papers) in the intersection of reproductive health and HCI. Their presence appears primarily in three roles: Fertility challenges (e.g, \cite{32,33,34,35}), Supportive Partners (e.g., \cite{83, 85}), and Decision-Makers in patriarchal contexts \cite{80}. While some interventions engage men constructively, others reinforce traditional hierarchies or ignore their emotional and reproductive needs entirely. 

\subsubsection{Marginalized Groups} Reproductive injustices experienced by minority groups, such as, migrant and refugee women, disabled or chronically ill women, or LGBTQ+ Communities are documented in 16 papers (e.g., \cite{21,69,79,128}). Despite these efforts, intersectional populations remain underrepresented in both study designs and technology development processes.

\subsubsection{Researchers/Designers} Researchers, designers, and practitioners of reproductive health and bodily experiences in HCI, critique reproductive health technologies to discuss, who benefits from these technologies, who is excluded, how do these technologies reinforce binary understandings of gender and fertility, and if we can reimagine reproductive health technologies beyond tracking and diagnosis (e.g., \cite{1, 22, 65}). Some researchers critique the narrow framing of “women's health” in HCI. Some studies highlight the lack of inclusion of trans, disabled, racially marginalized, and socioeconomically diverse groups in design and participatory design.

While women, healthcare providers, and researchers dominate the eco-system, two pivotal groups shape the systemic outcomes, namely, the digital platforms that mediate health technologies and the policymakers or advocacy groups that shape reproductive health discourse and governance. Post-Roe, many menstrual apps (e.g., Flo, Clue) were found to share sensitive user data, such as, GPS locations and cycle patterns with third-party advertisers (e.g., \cite{39,52,53,55,56,57}) Some apps did not provide any privacy notice upon installation or later use, and gave users little to no control over their personal data while using deceptive consent mechanisms (e.g, \cite{15,18,44,48,52,53,55,56,57}. AI-powered fertility apps disproportionately cater to Western users, often failing to reflect the lived experiences of women from other parts of the world. Western users, which may fail to reflect women's experiences from the rest of the world (e.g., \cite{26, 91, 134}). Platform developers are rarely held accountable for these harms, and the underlying technologies often prioritize profitability, engagement metrics, and investor returns over safety, inclusivity, or informed consent. This commercialization of care raises concerns about technological paternalism and the depoliticization of intimate health data. 

\subsection{Why: Focus Areas}

The reviewed literature on reproductive health and HCI reflects a wide range of motivations that have shaped the field over the past decade. Early work centered on addressing stigma and gaps in reproductive care, particularly in low-income regions or for marginalized populations, while most recent studies have focused on design justice, digital privacy, and participatory empowerment. Below, we organize the reviewed literature into key focus areas, highlighting how each contributes to reshaping reproductive health.

\subsubsection{Destigmatizing Reproductive Health} Stigma around reproductive health is particularly pronounced in low-income and resource-constrained settings where social taboos, limited access to care, and cultural silences often prevent individuals from discussing menstruation, infertility, or pregnancy loss. Studies in this context reveal how fear of judgement, religious or familial expectations along with lack of privacy can lead to emotional isolation or delayed care-seeking \cite{3, 14, 17, 47, 113, 122}. In response to these barriers, many studies prioritize destigmatization as a central concern, seeking to challenge shame and silence across a range of experiences including menstruation, infertility, abortion, pregnancy loss, and menopause. For example, Mustafa et al. reveals how purdah (modesty), and haya (shame) prevent open discussion of menstruation and menopause \cite{3}, causing many to rely on informal networks due to stigma around reproductive care. Analysis of menstrual health education (MHE) shows schools avoid direct terms (e.g., “uterus”), using euphemisms like  “monthly cycle”, reinforcing shame. Rural parents oppose MHE, fearing it “corrupts” children \cite{17}. Additionally, pregnant women were hesitant to disclose their pregnancies publicly due to shame and social stigma \cite{78}. Fear of being judged, gossiped about, or stigmatized discouraged open storytelling in some contexts \cite{82}. 

However, many studies highlighted how online forums offered safe spaces for individuals experiencing infertility or miscarriage to share personal narratives, seek community support, and push back against prevailing stigmas  \cite{32, 35, 105}. Building on these insights, later work has focused on designing systems or digital tools that not only support disclosure but actively challenge stigmas around reproductive health by introducing culturally resonant, activist-oriented interventions. For example, Michie et al. engaged pro-choice stakeholders through conducting interactive design workshops which enabled the participants to share experiences and co-develop counter-narratives against dominant societal stigma \cite{13}. Similarly, other studies employ zines, wearables, or chatbots to reclaim reproductive agency through visual and digital storytelling \cite{6, 42, 82}. Importantly, some studies addressed the stigma faced by men, showing how reproductive health framed exclusively as a “women's issue” leads to exclusion and emotional silencing among male users \cite{31, 32, 33, 34, 35, 37, 40}.  Together, these efforts reflect a sustained effort in reshaping reproductive care in both online and offline spaces.

\subsubsection{Promoting Health Literacy and Education} Promoting reproductive health literacy is a crucial focus across the reviewed studies, particularly in addressing the informational gaps and misinformation that disproportionately affect adolescents, low-literacy users, and marginalized communities. Accessible, engaging, and culturally grounded education tools are essential to enable informed reproductive-decision making \cite{10,12, 17, 34, 92, 104}. For example, Davinny Sou et al. employed surveys to assess menstrual health knowledge and introduced a menstrual health tracking app to improve awareness and literacy \cite{10}. Wood et al. used participatory body mapping and Lego-based activities to co-design educational tools with adolescents \cite{12}. Similarly, Gupta et al. introduced a game called “SheHealthy” using contextual interviews with girls, gynecologists, and health workers to design an interactive platform that communicated reproductive health knowledge.  A. Tuli et al. revealed gender-based disparities around menstrual health education, while Robins et al. analyzed the readability of male fertility websites, finding that many popular sources were not accessible to average readers \cite{17, 34}. Kumar et al. took a multidimensional approach to evaluate culturally specific education by reviewing menstrual comics, community video campaigns, and survey data in India \cite{92}. These examples reflect an expanding commitment to not only provide information, but to deliver it in forms that are linguistically, visually, and culturally resonant.
    
Consequently, these studies indicate that reproductive health literacy is not just a content problem, but a design challenge as it demands culturally situated strategies, inclusive media formats, and co-created knowledge-sharing tools to ensure reproductive autonomy for diverse user groups.

\subsubsection{Expanding Access to Reproductive Care} Expanding access to reproductive care, particularly for marginalized, low-income, or underserved communities, is a recurring and urgent focus across the reviewed studies. Researchers highlight that the lack of access to reproductive care is often shaped by limited infrastructure, language mismatches, and systemic exclusion from formal health systems. Consequently, to bridge these gaps digital tools are designed and proposed as equity-driven interventions. For example, some studies reveal how Mobile health interventions are used to deliver maternal health and prenatal care information in accessible formats \cite{72, 73, 74}. Similarly, Wang et al. evaluates a bidirectional SMS platform to deliver maternal health and family planning information timely \cite{41}. These approaches prioritize simplicity, affordability, and use of local language to ensure uptake and usability among populations with low literacy or digital access. These studies frame these limitations not only as infrastructural failures but as design challenges which require rethinking how care is delivered and who it is designed for \cite{22}.

\subsubsection{Safeguarding Privacy and Autonomy} As reproductive health applications become more commonly used, safeguarding privacy and reinforcing autonomy has become a priority. Studies highlight how Femtech platforms, specifically, period and fertility tracking apps often fail to meet ethical standards for data protection, exposing users to surveillance \cite{15, 30, 43, 52, 53, 55, 56, 57}. In the post-Roe context, it was found that logging data on these apps could reveal highly sensitive information, such as potential abortion timelines or clinic visits \cite{125}. Furthermore, Dong et al. reveal how some apps transmitted users' period-related health data (like cycle start/end dates, symptoms, mood, and sexual activity) to remote servers \cite{53}. Inconsistencies between privacy policies and in-app experiences undermine user trust \cite{44}.  To mitigate these risks, several studies call for privacy-by-design frameworks, transparent data governance, and culturally sensitive consent flows \cite{46, 56, 57}. 

\subsubsection{AI Ethics and Trust in Technology} Discussions surrounding the ethical implications of AI and algorithmic decision-making in reproductive health contexts are increasingly gaining attention. These studies interrogate how AI systems, particularly in fertility prediction and reproductive tracking, present themselves as authoritative, despite being trained on narrow datasets and offering limited transparency \cite{38, 134}. This misplaced trust becomes problematic in sensitive domains like conception planning, where emotional and clinical stakes are high. Some apps provide algorithmic fertility predictions that can sometimes be scientifically accurate to users, even when models rely on generalized assumptions \cite{26, 43}. This misplaced trust becomes problematic in sensitive domains, such as, conception planning, where emotional and clinical stakes are high \cite{11}. 
    
Other studies point to the transparency of AI models, as some reproductive platforms provide vague or evasive language around algorithmic operations, in response to the overturn of Roe v. Wade \cite{61}. Users feel disempowered when they are given little to no control over their data, and distrust towards these apps increases \cite{46}.  Additionally, trusting these tools gets complicated because users do not have enough information and face language and cultural barriers \cite{75}.  Consequently, the question arises: what constitutes trustworthy AI in reproductive health? This shifts the focus from predictive accuracy to accountability, inclusivity, and ethical design practices.

\subsubsection{Addressing Men's Reproductive Health} Most reproductive health research still focuses on women, but some fewer studies are starting to look at men's experiences–specifically with infertility, emotions, and fatherhood. These studies aim to reveal the often-silenced emotional and relational burdens faced by men while challenging the notion that reproductive health is exclusively a "women's issue." A lot of this research focuses on infertility and the psychological distress it causes, and documents how men navigate feelings of shame and loss of masculinity. For example, anonymous online infertility forums give men a space to express grief, seek advice, and connect with others facing similar struggles \cite{32, 35}. While men recognize the importance of preconception health, their knowledge and access to resources are limited. Most fertility apps focus on women, highlighting the lack of male-specific digital health interventions \cite{33}. These studies address a significant gap in reproductive health research and advocate for design frameworks that recognize men as caregivers, patients, and stakeholders in reproductive ecosystems.

\subsection{How: Research Methods}
The reviewed papers employ a wide range of research methods, including qualitative, quantitative, and mixed approaches, participatory and design-led methods, offering a diverse perspective on the complexities, barriers, and possibilities within reproductive care. Below, we outline the primary methodological approaches.

\subsubsection{Qualitative Approaches} A majority of the papers adopt qualitative methods to capture the lived experiences of the target participants. Techniques include interviews, focus group discussions, ethnography, discourse analysis, and content analysis of online forums and apps (e.g., \cite{58, 64, 102, 122}).  Studies often use thematic analysis (e.g., \cite{7, 10, 12, 17, 32, 36, 40, 43, 48, 50, 58, 60}) or reflexive thematic analysis (e.g., \cite{48, 58, 120}), grounded theory (e.g., \cite{66, 85, 96, 107, 122}), and critical discourse analysis \cite{2} to investigate meanings, emotions, and power dynamics embedded. For example, Mustafa et al. interviewed Muslim women in Bangladesh, Pakistan, and Malaysia to explore how religious values influence navigation around reproductive health \cite{3}. Andalibi interviewed participants to understand and analyze disclosure practices after pregnancy loss \cite{66}. Additionally, African-American women's reproductive health information-seeking behaviors \cite{102} and Arab Muslims' intimate health literacy \cite{86} were also explored through interviews.
    
Ethnographic studies offer insights across time. Zhao et al. uses a two-year ethnography with migrant women in Shenzhen to document the intersections of marriage, labor, and reproductive precarity \cite{122}. Digital ethnography (e.g., Reddit, Twitter, BabyCenter)  also plays an important role, particularly studies analyzing, miscarriage, menopause, or online support systems (e.g., \cite{32, 105, 107, 117}).  Several papers use feminist paradigms, emphasizing researcher's standpoint reflexivity and positionality (e.g., \cite{2, 21, 84,87, 89, 95}). Some papers incorporate visual ethnography (e.g., \cite{78}), body mapping (e.g., \cite{12, 20, 78}), and sensory bodystorming (e.g., \cite{54}), expanding the qualitative approach beyond textual data. 

\subsubsection{Quantitative Approaches} Multiple studies used quantitative approaches to analyze large-scale trends, app privacy practices, user behaviors, and policy gaps. These include online surveys (e.g., \cite{10, 23, 31, 34}), RCTs (e.g., \cite{41, 74, 80}), and controlled experiments (e.g, \cite{38, 46}). Costa Figueiredo et al. tested an app environment to understand how different AI descriptions impact trust in fertility apps, using ordinal regression and Kruskal-Wallis tests \cite{38}. Some studies conduct app audits to quantify data-sharing violations in 30 menstrual trackers \cite{57} or compare GDPR compliance across 85 fertility websites \cite{34}. 

\subsubsection{Participatory and Design-Led Methods} Many papers engage in participatory design (PD), research-through-design (RtD), or speculative design to co-create technologies with users (e.g., \cite{6, 12, 13, 21, 36, 84, 88, 99, 116}). These methods are grounded in feminist HCI, intersectional theory, and design justice frameworks (e.g., \cite{9, 17, 19,  22, 26, 42, 46, 48, 67, 72, 81, 91, 92}). For example, South-African midwives participate in co-designing the SHiMA maternal health app using CommCare \cite{88}. Co-design workshops, zine-making, scenario building, and prototype iteration are also some of the techniques employed (e.g., \cite{42, 65, 99, 103}). Some studies integrate RtD with long-term field study, while some experiment with metaphor-driven prototyping \cite{21, 116}.
    
\subsubsection{Mixed-Methods and Longitudinal Studies} A large number of papers employ mixed-methods approaches that combine surveys, interviews, app reviews, participatory tools, and statistical modeling. These methods offer complementary perspectives that blend scale with depth. For example, Ibrahim et.al conduct a mixed-method approach in three phases: two surveys and interviews to investigate menstrual tracking privacy post-Roe. Additionally, convergent mixed-methods combine surveys with interviews to evaluate trust in fertility apps \cite{38} or map FemTech users' privacy mental models via sketches and stories \cite{124}. Some studies combine qualitative thematic analysis with NLP or sentiment analysis of large datasets ( app reviews or social media posts) (e.g., \cite{45, 93, 125}). Mehrnezhad and Almeida combines surveys, story completion tasks, and sketch-based interviews to explore FemTech mental models of privacy \cite{124}.

\subsubsection{Longitudinal and Review-Based Studies} Longitudinal Studies are often qualitative and provide insights into behavioral changes, emotional trajectories, or system engagement over time. Almeida et al.'s eight-year RtD program, while Prabhakar et al. track low-income mothers and postpartum Facebook users over months-long periods \cite{85, 90}. Some studies track devices for months or engage in platform ethnography \cite{50, 108}. Several papers adopt systematic or scoping reviews to explore trends across reproductive health domains \cite{8, 27, 52, 112}. Lastly, a few papers propose frameworks based on secondary analysis or citation analysis, for example, Chivukula analyzes 70 HCI citations of Bardzell (2010) to evaluate feminist uptake in the field, while Fledderjohann, Knowles \& Miller built a reproductive justice-informed critique of algorithmic harms \cite{95, 119}.

%% file: 3.tab-3.tex
\begin{table}[!t]
\centering
\begin{tabular}{|p{1.4cm}|p{5.75cm}|p{5.75cm}|}
\hline
\textbf{Waves} & \textbf{Global North} & \textbf{Global South} \\
\hline

\textbf{First Wave:} &
\parbox[t]{5.75cm}{
(1) Uses advanced apps and wearables with interactive, user-centered features.\\
(2) Focuses on psychological well-being, peer support, and integration with healthcare systems.\\
(3) Strong online communities supported by high digital access and literacy.}
&
\parbox[t]{5.75cm}{
(1) Relies on low-cost, accessible communication methods (SMS, voice calls, and basic mobile phone functions) suited for resource-constrained settings with limited infrastructure or digital literacy.\\
(2) Focus on basic access to information and care in underserved areas.\\
(3) Peer support is limited but gradually emerging. \\}
\\
\hline

\textbf{Second Wave:} &
\parbox[t]{5.75cm}{
(1) Strong focus on privacy, data security, and anonymous user interaction.\\
(2) Advocates to align technologies with societal norms, such as designing apps to be discreet or gender-neutral to reduce stigma, including local languages and norms, ensuring relevance and user trust.\\
(3) Emphasis on inclusive, feminist design to avoid stereotypes and exclusion.}
&
\parbox[t]{5.75cm}{
(1) Low-tech solutions are designed for affordability and accessibility to include low-income and low-literate users and help deliver health information where internet access is limited. These solutions overcome infrastructural challenges, such as unreliable electricity or internet, ensuring basic maternal health information reaches users.\\
(2) Content is localized and spread through community workers to reach rural and marginalized groups. \\}
\\
\hline

\textbf{Third Wave:} &
\parbox[t]{5.75cm}{
(1) AI integration in FemTech.\\
(2) Algorithmic bias is a critical concern in FemTech, prompting calls for a participatory audit.\\
(3) After Roe v. Wade was overturned, users in the Global North turned to VPNs and encryption to protect their sensitive health data, as trust in app privacy dropped.\\
(4) Critical design response to challenge gender and fertility biases by using inclusive language and involving marginalized users in the design process. \\}
&
\parbox[t]{5.75cm}{
(1) Uses low-tech tools like SMS, chatbots, and local videos to improve reproductive health access and education, while calling for inclusive AI and participatory design to reduce bias.\\
(2) Focuses on reaching marginalized users, improving menstrual literacy, and designing tools that respect local cultures, faiths, and languages through community-based approaches.}
\\
\hline
\end{tabular}
\caption{Reproductive Well-being Research Themes in Three Waves between Global North and Global South}
\label{tab:repro_wave}
\end{table}

%% file: 5.insights.tex
\section{Insights into what is missing}
Despite expanding significantly, the reproductive well-being scholarship in HCI and related domains has remained constrained by narrowing its focus to the needs and concerns of cisgender women and overlooking broader stakeholder roles, regional differences, and alternative temporalities. This section identifies the major gaps in existing reproductive well-being scholarship in HCI and related domains in three major themes: absence of other stakeholders, missing contextual mapping across geographies, and poor recognition of alternative temporalities.

\subsection{Significant Absence of Other Stakeholders}

Research regarding reproductive well-being has historically centered cisgender women as the default users, reinforcing gendered assumptions about who bears responsibility for reproductive well-being \cite{2}. While this led to valuable tools like menstrual and pregnancy tracking apps, it has also created blind spots by excluding other critical actors, such as men, non-binary individuals, and caregivers like parents, in-laws, and community health workers. 

Men's reproductive health, in particular, is significantly underexplored. Throughout the literature, there are very few tools designed for men’s fertility, emotional support during pregnancy, or shared roles in contraception \cite{32,33,34,35}). Men are often mentioned in passive or secondary roles, such as partners or decision influencers, rather than as individuals with reproductive needs and experiences of their own. This narrow framing limits opportunities for shared responsibility and overlooks how partners contribute to reproductive wellbeing within families and communities, while reinforcing the stereotype that reproduction is a "women's issue." Similarly, adolescents are less represented in the field despite going through critical stages of sexual development. LGBTQ+ users also face exclusion, as most technologies assume binary gender roles and heterosexual family models. Beyond individuals, informal caregivers like parents, in-laws, and family members and community health workers are often central to reproductive decision-making in many cultural settings, but they are rarely included in the design process. Migrants, refugees, and people with disabilities, who face systemic barriers to access and support, are similarly underrepresented. Consequently, this pattern reflects a larger design gap: when technologies ignore the full ecosystem of stakeholders, they fail to serve the broader, real-world dynamics of reproductive well-being.

\subsection{Missing Contextual Mapping Across Geographies}

A major gap in reproductive well-being is the lack of contextual adaptation to the diverse cultural, religious, and policy environments across different regions. Many solutions for reproductive wellbeing are developed with a one-size-fits-all approach, which reflects the Western norms and values, and may not align with the needs and realities of users in other parts of the world. By “context”, we refer to the specific cultural, religious, policy-based, and intervention-driven factors that shape reproductive health experiences in a given region, including local norms, government or NGO programs, healthcare infrastructure, and national reproductive policies.

Cultural beliefs and religious practices play an important role in shaping how reproductive health is perceived and accessed. These perspectives differ across global regions, including parts of Asia, North America, Europe, and Africa. For instance, in Bangladesh, prevailing patriarchal norms and a cultural preference for sons result in early marriages (often around the age of 14.5 years), and higher fertility rates, despite economic challenges \cite{sultana2018design}. These societal expectations often confine women to domestic roles, limiting their autonomy and access to education and healthcare. Community health workers or their local healthcare providers frequently serve as the primary source of reproductive health information and services for these women \cite{97, 135}. Religious norms further shape the perspectives towards reproductive wellbeing. For instance, Islamic values in Arab regions demand culturally sensitive health education, as seen with users seeking Islamic-aligned content \cite{7, 86}. Additionally, national policies regarding reproductive health can differ dramatically, and often define the legal and logistical boundaries of reproductive care. Government and NGO implemented programs also shape the framing of reproductive well-being. In regions with weak public health infrastructure, NGOs often lead in providing reproductive education and services \cite{78, 97}. Healthcare infrastructure also determines access to care, especially in low-resource settings where overcrowding, language gaps, and mistrust make formal services hard to use \cite{135}.

However, how well these interventions work depends on whether they fit local culture and meet the needs of the people they aim to support. Without sufficient contextual mapping, research and interventions risk applying generalized solutions to particular needs. This can result in low adoption, user distrust, and unaddressed disparities in care. For reproductive well-being research to be meaningful and impactful, it needs to be rooted in specific regional contexts and shaped by the voices of local communities to reflect their real-life needs and experiences.

\subsection{Poor Recognition of Alternative Temporalities}
\textbf{\textit{Alternative temporalities}} refer to ways of experiencing, organizing, or understanding time that differ from dominant, linear, Western notions of time (e.g., clock time, progress, productivity) \cite{freeman2010time, bhabha1994location, lefebvre2004rhythmanalysis, bertens2021alternative}. The concept is widely used in critical theory, cultural studies, and feminist and postcolonial scholarship to describe temporal frameworks that emerge from different social, cultural, or embodied experiences. Linear and non-linear time are two of the core key concepts in temporality. In \textbf{linear time} sensitivities, time progresses forward in a straight, measurable, and irreversible path—past to present to future. Linear time is characterized by chronological sequencing, a focus on causality and progress, reliance on clock or calendar-based measurement, and dominance within industrial, capitalist, and Western epistemologies. On the other hand, \textbf{non-linear time} sensitivities state that time is experienced as cyclical, recursive, fragmented, or layered rather than progressing in a straight line. Non-linear time is marked by cycles (e.g., menstrual cycles, seasons, rituals), recurrence and return (e.g., trauma, memory), multitemporal coexistence (e.g., past, present, future folded into the now), and flexible or suspended timelines (e.g., waiting, interruption, chronicity).

Most of today's tracker tools of reproductive well-being embrace linear sensitivities of time instead of non-linear sensitivities. This dismissal of alternative \textbf{calendars}, \textbf{cycles}, and \textbf{body-clocks} is particularly problematic for the users and target population that are not "standard" in the perspectives of the tracker designers, as we found evidence in the literature. Many users—especially those from marginalized or culturally specific contexts—experience a fundamental dissonance between their lived temporalities and the linear assumptions embedded in these technologies \cite{14, 18}. Many existing menstrual tracking tools adhere to linear, clock-based models that fail to account for alternative calendars, such as religious calendars or culturally defined cycles, leading to user frustration when trying to align health practices with ritual observances \cite{9}. 

These design gaps stem from a dominant focus on \textbf{linear reproductive timelines} that centers on ovulation, fertility windows, or conception goals and overlooks the complex, non-linear rhythms of bodies and lives. Users navigating irregular cycles, chronic conditions like polycystic ovary syndrome (PCOS), or culturally situated reproductive practices require tools that embrace flexible, cyclical, and phase-based tracking \cite{14, 18}. Yet most applications continue to prioritize TTC (trying to conceive) logics, underserving life-stage transitions such as menopause or postpartum recovery, and neglecting those avoiding pregnancy (TTA) or not conforming to normative reproductive paths \cite{18, 28}. This failure to reflect the diverse, recursive, and nonlinear realities of the users is not only a failure of the tools to serve the users, but also deceptive and causes frustration and trauma among the users who have limited access to resources \cite{9}.

%% file: 6.discussion.tex
\section{Discussion}
This paper maps the literature on reproductive well-being in the past ten years and finds that reproductive well-being technologies are never neutral; rather, they reflect embedded values, normative assumptions, and sociopolitical conditions that shape who is supported, excluded, or surveilled. Our review also reveals that while digital tools have expanded access and visibility, they often reinforce gendered burdens, cultural mismatches, and infrastructural inequities. Below, we discuss the broader implications of our findings and propose the Reproductive Well-being for All (ReWA) framework. 

\subsection{Broaders Implications}
\subsubsection{Cultural Norms, Stigma, and Institutional Gaps}
While most reproductive well-being technologies are expected to be supportive of users' reproductive well-being, in practice, we found many of them failed to cater to the needs of the users, particularly those who are in complex socio-cultural settings \cite{3, 5, 140, 148}. Additionally, we noticed that reproductive well-being technologies built based on modern scientific knowledge set and rationality can potentially reinforce existing structural inequalities through norms, stigma, and institutional neglect \cite{3, 92, 108, 140}. For example, talking openly about reproductive and sexual health is still taboo in many societies, which makes it hard for people to speak up or get support \cite{3, 5, 63, 135}. In conservative or faith-based social settings, cultural norms of modesty restrict conversations about menstruation, contraception, and sexual wellbeing \cite{17, 70, 78, 79, 132}. For example, Muslim women align their reproductive behavior, such as menstrual tracking, with religious practices like fasting, prayer, and purification \cite{7, 9}. However, these intentions often clash with cultural taboos that make it hard to talk openly, treating menstruation and menopause as shameful \cite{3, 17, 28, 91, 105, 109}. Additionally, cultural myths continue to dominate menstrual health education and practices, such as women may be kept out of places like temples or kitchens during their menstruation \cite{17}. Menopause is also rarely talked about, and cultural ideas about aging and femininity often lead to shame and confusion \cite{28}. Educational and community-awareness driving institutions' silence makes it worse, as many teachers avoid or quickly go over menstrual health and reproductive hygiene topics because of cultural taboos and personal discomfort, especially male teachers and teachers in mixed-gender classes in the Global South \cite{17}. 


These cultural and institutional dynamics have been reported to directly affect how people experience and respond to technology \cite{3, 17, 42, 64, 91, 109}. Whether it is quiet, private talks about periods or pushback against menstrual cups because of purity beliefs, social and cultural norms shape what health information people accept and which technologies they trust \cite{67, 70}. In places where people are less likely to trust formal systems, they often use informal networks like WhatsApp to talk about reproductive issues in a safe and private way \cite{63, 79, 109, 130}. These spaces are not just created out of necessity but out of resistance to systems that have historically failed to listen \cite{13, 42, 89, 106, 130}. At the same time, people dealing with sensitive issues like pregnancy loss or infertility often feel emotionally isolated, which is made worse by a lack of support and silence in their communities \cite{66}. While it is clear that most of today's theories, practices, and designs of reproductive well-being tools and technologies fail to address these challenges, not only that, but how such challenges, stigma, cultural taboos, and weak institutional acts reinforce the challenges in different parts of the world is still understudied in HCI and related domains. This gap in knowledge makes it harder to respond to the social and emotional needs of reproductive well-being of people across the world. Such gaps in knowledge can be addressed by cross-talks between the domains, such as active knowledge sharing among domains of HCI, AI, social computing, medicine, medical anthropology, statistics, and NGO and local governments. However, researchers in several of the domains mentioned above would still be required to be open to \textbf{methodological pluralism} and sensitive to \textbf{contextuality} to understand the nuances beyond scientific rationality and embrace alternative rationality, as several well-being research from the global south suggests \cite{sultana2019witchcraft, sultana2021dissemination, sultana2019parar, sultana2020parareligious, sultana2021opaque, sultana2020fighting, mahzabin2024ancient}. 


\subsubsection{Technology as both a Tool and a Barrier}
Reproductive technologies offer new opportunities for self-tracking, education, and care, but they also reveal deep structural limitations shaped by cultural assumptions, data practices, and social norms. Despite the growth of apps for menstruation, fertility, and sexual health, many users—especially in the Global South—report a mismatch between their lived experiences, locality and practicing norms,  and the design logics of these tools and cited concerns over judgment, low literacy, and the absence of culturally sensitive interfaces \cite{3}. For example, many apps fail to accommodate religious calendars, frustrating users who wish to align health tracking with spiritual routines \cite{9}. Design gaps persist as most tools are built around narrow parameters like ovulation windows or bleeding days, excluding users with non-normative cycles, PCOS, or those navigating menopause or postpartum life stages \cite{14, 18, 28}. In high-surveillance contexts like the post-Roe United States, period tracking apps have raised alarm over data misuse, with reports of intimate information, like location or sexual activity, being transmitted without justification \cite{43, 53, 55}. These fears reveal how technologies can be weaponized, undermining both trust and safety. Still, users continue to adapt and repurpose these tools, resisting their limitations. Technologies for reproductive well-being are not neutral—they reflect embedded values that determine whose needs are prioritized and which bodies are considered ``normal." HCI-design scholarship's responsibility here is more than just UX fixes; it demands a critical rethinking of the cultural, ethical, local history, and political assumptions underpinning these systems.

\subsubsection{Gendered Burdens and Responsibilities}
Despite advancements in reproductive technologies, our review reveals that the emotional and mental labor of managing reproductive well-being continues to fall disproportionately on women and menstruators. Across platforms and daily practices, women are framed as the primary caregivers and information managers—expected to track fertility, monitor symptoms, and interpret health data often without shared support \cite{4, 11, 14, 26}. This responsibility extends beyond tasks to emotional labor: users describe mixed feelings of empowerment and anxiety, with apps sometimes exacerbating stress or leading to burnout. Digital tools offer little support for navigating complex experiences like PCOS, pregnancy loss, or menopause, leaving many women feeling isolated and overwhelmed. Although peer networks sometimes provide comfort, the systemic lack of shared responsibility from partners, families, and institutions persists. Male partners are typically depicted in passive roles, if at all, reinforcing the idea that reproductive care is a woman’s domain and leaving women to manage uncertainty and disappointment alone \cite{17, 28}.

Yet men are not untouched by reproductive burdens; rather, their challenges are often overlooked or silenced in the literature. Men are expected to possess reproductive knowledge without adequate education or safe spaces for dialogue \cite{31}. Infertility, in particular, carries a deep emotional toll—men report feelings of guilt, inadequacy, and emasculation, especially when their partners undergo invasive treatments for male-factor infertility \cite{32, 35}. Such silencing and emotional tolls on men are harmful for both the men and the people around them, and this is against feminist well-being agendas. Online forums and digital communities have become critical outlets, helping men share emotions and disrupt norms that demand silence or stoicism \cite{32, 33, 35, 37}. However, gender inequality also shapes who controls technology and reproductive decisions: in many households, men dominate device access and information flows, while women may resort to using SMS-based programs in secret \cite{sultana2018design, 41}. These dynamics create asymmetrical access and reinforce cultural narratives that place visible blame on women while internalizing shame in men. 

Building on recent calls in HCI, social computing, design, and ICTD for "gender de-stereotyping" of reproductive care by expanding beyond women-centered designs \cite{almeling2013more, kukura2022reconceiving, 135}, we argue that these gendered divisions call for a shift in how we conceptualize and design reproductive well-being technologies. Reproductive care should not be treated as a women-only issue, it is a shared human concern that affects individuals across genders, relationships, and life stages. Centering only women in both design and discussion not only reinforces inequalities but also overlooks the needs, voices, and experiences of men and non-binary individuals. To support reproductive well-being for all, technologies must move beyond gendered assumptions and embrace a model that supports collective care, encourages open communication between partners, and addresses the diverse realities of users. For HCI and design research, this calls for reframing reproductive well-being as a shared, relational concern, demanding tools that redistribute cognitive and emotional labor, enable co-participation, and challenge entrenched gender norms.

\subsubsection{Polyvocality and Inclusive Storytelling}
Reproductive technologies often follow a singular, linear script—menstruation to conception—excluding diverse narratives like miscarriage, assisted conception, non-conception by choice, or queer and menopausal experiences \cite{26}. Yet across the studies we reviewed, users pushed back through storytelling, humor, and peer-to-peer knowledge sharing, shifting from passive data points to active narrators of their reproductive lives. Tools like \textit{Labella} and \textit{Menoparty} created safe spaces for reflection and emotional expression, allowing discomfort, grief, and humor to coexist \cite{25, 28}. On the other hand, platforms like Facebook were seen as too polished for expressing pregnancy loss, while WhatsApp groups, zines, and professional-run Instagram accounts offered more intimate and culturally grounded storytelling spaces \cite{47, 63, 64}. These informal networks often addressed gaps left by formal care systems, enabling users—especially those from marginalized backgrounds—to share context-specific knowledge shaped by religion, identity, and lived experience. In some cases, storytelling became an act of resistance: men submitted false period data to confuse surveillance systems \cite{43}, and used memes to express reproductive concerns with emotional nuance \cite{31}. Designs that embraced real-life messiness—rather than only clinical metrics—were more inclusive. Feminist disability studies and soma design approaches helped frame uncertainty, tension, and shared pain as legitimate parts of reproductive experience \cite{42, 54}. Ultimately, storytelling enables reproductive justice by making space for complex, open-ended, and non-linear experiences. HCI and design research should emphasize designs that value plural voices, recognize narrative agency, and support not just what can be measured, but what must be told.

\subsubsection{Agency in Action: How Users Are Shaping Reproductive Well-Being Design}
Across reproductive technologies, users are asserting themselves not just as patients but as critics, co-creators, and change-makers. Through design critiques, community feedback, and digital activism, they are challenging exclusion, misinformation, and surveillance. For example, users pushed back against privacy violations by deleting apps, demanding encryption, and calling for open-source alternatives \cite{43}, with many placing greater trust in app developers than governments to safeguard their data \cite{46}. Additionally, community platforms have become grassroots design spaces—Reddit threads offer support for managing PCOS \cite{14}, while WhatsApp groups create trusted circles for discussing menstruation, sexuality, and emotional well-being \cite{63}. Users actively call for greater transparency and consent, especially after studies revealed that many menstrual apps failed to comply with GDPR and collected excessive personal data \cite{52, 53}. Male users, often excluded from mainstream design, expressed clear needs for culturally sensitive, accessible tools for preconception care \cite{36, 39, 41}. This growing user agency signals a shift: people now evaluate reproductive technologies not only by functionality but by whether they foster trust, inclusion, and justice. For HCI and design research, this means reimagining users as collaborators, whose lived insights, resistance, and demands are vital to shaping equitable reproductive futures.

\subsection{Toward \textit{Reproductive Well-being for All (ReWA)}}

\begin{wrapfigure}{r}{0.5\textwidth}
\vspace{-18pt}
\includegraphics[width=0.99\linewidth]{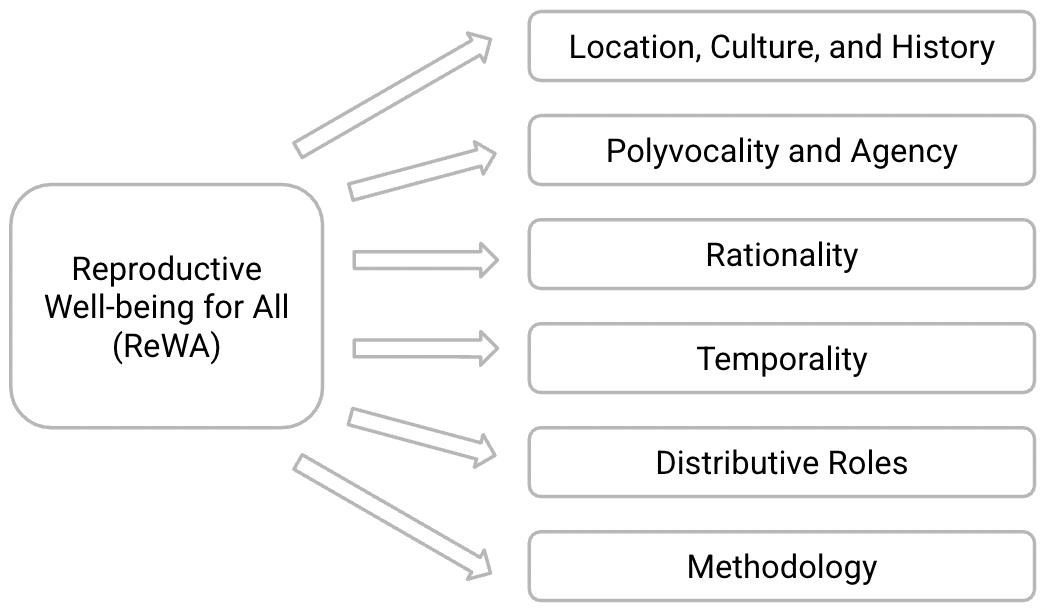} 
\caption{The \textit{Reproductive Well-being for All (ReWA)} framework consisting of six dimensions: location, culture, and history; polyvocality and agency; rationality; temporality; distributive roles; and methodology.}
\label{fig:wrapfig}
\vspace{-15pt}
\end{wrapfigure}

Drawing on the existing literature, we propose the \textbf{\textit{Reproductive Well-being for All (ReWA)}} framework. This framework will work as a set of practical guidelines to design fair, accountable, and sustainable infrastructure and computing tools to support reproductive well-being for all, as well as a critical lens to analyze whether an existing infrastructure or tool is supportive of reproductive well-being for all. This framework consists of six orientations associated with the following aspects: location, culture, and history; polyvocality and agency; rationality, temporality, distributive roles, and methodology. We detail them below:

\subsubsection{Location, Culture, and History}
Reproductive well-being support tools must account for the politics of location, culture, and history embedded in users' data. As reproductive experiences and meanings shift across contexts, such as migration, faith, or stigma, systems need epistemologically rich, situated approaches to interpret them responsibly. Yet current infrastructures and tools often fail to recognize how the value and interpretation of reproductive data change across contexts, revealing a core limitation in dominant HCI, social computing, and AI design paradigms.

\subsubsection{Polyvocality and Agency}
To support reproductive well-being for all, infrastructure, tools, and technologies need to embrace polyvocality by amplifying the diverse stories, values, and epistemologies that shape people's reproductive lives. This includes acknowledging religious, experiential, and community-based plural knowledge systems as valid alongside scientific and medical frameworks, and creating space for users to annotate, contest, or co-author their data narratives. Without mechanisms for expression, resistance, or refusal, reproductive technologies risk flattening complex realities into narrow logics of productivity, risk, or compliance.

\subsubsection{Rationality}
Dominant reproductive technologies often rely on Western scientific, medical, and algorithmic rationality approaches that emphasize quantification, risk prediction, and standardized outcomes. However, such frameworks can obscure or delegitimize other forms of knowing and rational decision-making that are deeply embedded in users' cultural, spiritual, emotional, and collective experiences, i.g, \textit{alternative rationalities} \cite{135, sultana2019witchcraft, sultana2021dissemination, sultana2020parareligious}. Reproductive technologies that ignore these forms of reasoning risk epistemic violence, rendering users' practices invisible or deviant. To design more usable-to-all systems, HCI and CSCW must expand their epistemic scope, honor and integrate plural rationalities, and create tools in design and practice that support without overwriting how people make sense of their reproductive lives.

\subsubsection{Temporality}
Designers and researchers of reproductive well-being need to move beyond linear, progress-driven models and instead center \textit{alternative temporalities} that reflect the cyclical, disrupted, and layered rhythms of reproductive life \cite{freeman2010time, bhabha1994location, lefebvre2004rhythmanalysis, bertens2021alternative}. Dominant tools often impose clock-based logics—ovulation cycles, conception windows, symptom logs—while ignoring how time is experienced through waiting, recurrence, and ritual, particularly among users with irregular cycles, chronic conditions, or religious observances. Recognizing non-linear time sensibilities allows technologies to better align with users' embodied and culturally situated experiences, enabling more inclusive, respectful, and responsive reproductive care.

\subsubsection{Distributive Roles}
Design and research in reproductive well-being must shift away from framing it as a "women-only" concern and instead recognize the distributed roles of men, family members, non-binary people, and community actors in shaping reproductive decisions and care. Across many cultural contexts, decisions about menstruation, fertility, pregnancy, and contraception are deeply influenced by spouses, in-laws, and even employers or religious leaders. Designing inclusive technologies and studies requires accounting for these relational dynamics, acknowledging that reproductive well-being is co-produced through social, familial, and institutional negotiations.

\subsubsection{Methodology}
Design and research in reproductive well-being must actively foster cross-talk among domains such as HCI, medicine, public health, anthropology, and feminist theory, and sectors such as local government and NGOs working on such projects, to capture the full complexity of users' experiences. Only cross-disciplinary and cross-sector movements can account for the intersecting cultural, emotional, technological, and political dimensions of reproductive care. Embracing methodological pluralism that combines clinical evidence, ethnographic insight, participatory design, critical theory, and mass-scale institutional evaluations is essential to building technologies that are both contextually grounded and socially accountable.

\subsection{Limitations of This Work}
While this review offers a comprehensive synthesis of reproductive well-being research across HCI, CSCW, and related fields, it is not without limitations. First, our selection was limited to English-language, peer-reviewed publications, which may exclude valuable insights from non-English, gray literature, or grassroots design practices published in local forums or non-academic venues. Second, although we included papers from a broad range of global regions, the literature itself remains disproportionately skewed toward studies conducted in the Global North, particularly the United States, limiting generalizability across diverse sociocultural contexts. Third, the review draws primarily from published findings and does not capture ongoing or unpublished work that may reflect more recent shifts in design practice or policy response. Finally, while our thematic framework aimed to surface structural patterns and gaps, we recognize that intersectional experiences, especially among disabled people of different gender spectrums, or Indigenous users, are complex and may resist neat categorization. These limitations call for more global, multilingual, and community-driven reviews in the future to ensure fuller representation of reproductive well-being realities.

\section{Conclusion}
This review underscores the urgent need to reimagine reproductive well-being technologies through more inclusive, culturally grounded, and ethically aware design practices. While digital interventions have expanded access, they often reproduce structural exclusions, center cisgender women, and overlook the emotional, cultural, and political dimensions of reproductive care. By tracing three waves of research over a decade, we show how technological innovation has outpaced critical reflection on who is served, who is excluded, and whose realities are rendered invisible. Addressing these blind spots requires a shift toward participatory, polyvocal, and justice-oriented approaches that engage diverse users, not only as data points but as co-designers of their care systems. Future work in HCI and CSCW must reckon with the limitations of dominant framings and move beyond tracking and diagnosis toward supporting autonomy, relational care, and collective well-being. Only then can reproductive technologies truly reflect the complexity of lived experiences and contribute meaningfully to equitable health futures.

%% file: 6.lim-fw-con.tex

\begin{acks}
Anonymized for review.
\end{acks}